\RequirePackage{fix-cm}
\documentclass[allclo, onecollarge, square, comma, sort&compress, natbib]{svjour2}
\bibpunct{[}{]}{;}{n}{}{,} 

\usepackage{amsmath}
\usepackage{amssymb}
\usepackage{color}
\usepackage{graphicx}
\usepackage{placeins}
\usepackage{supertabular}
\usepackage{xspace}

\usepackage[mathscr,scaled=1.15]{urwchancal}
\DeclareFontFamily{OT1}{pzc}{}
\DeclareFontShape{OT1}{pzc}{m}{it}%
{<-> s * [1.15] pzcmi7t}{}
\DeclareMathAlphabet{\mathpzc}{OT1}{pzc}{m}{it}

\definecolor{purple}{rgb}{0.5,0,0.5}
\definecolor{blue}{rgb}{0.0,0,0.9}
\definecolor{prdblue}{rgb}{0.133,0.118,0.498}
\usepackage[colorlinks=true, pdfstartview=FitV, linkcolor=prdblue, citecolor= prdblue, urlcolor=prdblue]{hyperref}

\journalname{Few-Body Systems}
\begin{document}

\title{Spectrum of light- and heavy-baryons
}

\titlerunning{Spectrum of light- and heavy-baryons}

\author{S.-X.~Qin
    \and
        C.\,D.~Roberts
    \and
        S.\,M.~Schmidt
}


\institute{Si-Xue Qin \at
            Department of Physics, Chongqing University, Chongqing 401331, P.\,R. China.\\
              \email{\href{mailto:sqin@cqu.edu.cn}{sqin@cqu.edu.cn}}\\[-1ex]
            $\,$\\Craig D.\ Roberts \at
            Argonne National Laboratory, Lemont, Illinois 60439, USA.\\
              \email{\href{mailto:CDRoberts.phy.anl.gov@gmail.com}{c.d.roberts@anl.gov}}\\[-1ex]
            $\,$\\Sebastian M. Schmidt \at
            Institute for Advanced Simulation, Forschungszentrum J\"ulich and JARA, D-52425 J\"ulich, Germany.\\
            \email{\href{mailto:s.schmidt@fz-juelich.de}{s.schmidt@fz-juelich.de}}
}

\date{Received:
28 January 2019
}

\maketitle

\begin{abstract}
A symmetry-preserving truncation of the strong-interaction bound-state equations is used to calculate the spectrum of ground-state $J=1/2^+$, $3/2^+$ $(qq^\prime q^{\prime\prime})$-baryons, where $q, q^\prime, q^{\prime\prime} \in \{u,d,s,c,b\}$, their first positive-parity excitations and parity partners.  Using two parameters, a description of the known spectrum of 39 such states is obtained, with a mean-absolute-relative-difference between calculation and experiment of $3.6(2.7)$\%.  From this foundation, the framework is subsequently used to predict the
masses of 90 states not yet seen empirically.
\end{abstract}

\section{Introduction}
\label{intro}
The Faddeev equation was introduced almost sixty years ago \cite{Faddeev:1960su}.  It treats the quantum mechanical problem of three-bodies interacting via pairwise potentials by reducing it to a sum of three terms, each of which describes a solvable scattering problem in distinct two-body subsystems.  The Faddeev formulation of that three-body problem has a unique solution.

An analogous approach to the three-valence-quark (baryon) bound-state problem in quantum chromodynamics (QCD) was explained in Refs.\,\cite{Cahill:1988dx, Burden:1988dt, Cahill:1988zi, Reinhardt:1989rw, Efimov:1990uz}.  In this case, owing to dynamical mass generation, expressed most simply in QCD's one-body Schwinger functions in the gauge \cite{Bowman:2004jm, Boucaud:2006if, Boucaud:2011ug, Ayala:2012pb, Aguilar:2012rz, Binosi:2014aea, Aguilar:2015bud, Binosi:2016wcx, Binosi:2016xxu, Binosi:2016nme, Rodriguez-Quintero:2018wma, Gao:2017uox} and matter sectors \cite{Lane:1974he, Politzer:1976tv, Bhagwat:2003vw, Bowman:2005vx, Bhagwat:2006tu, Chang:2010hb}, and the importance of symmetries \cite{Munczek:1994zz, Bender:1996bb, Binosi:2016rxz}, one requires a Poincar\'e-covariant quantum field theory generalisation of the Faddeev equation.  Like the Bethe-Salpeter equation for mesons, it is natural to consider such a Faddeev equation as one of the tower of QCD's Dyson-Schwinger equations (DSEs) \cite{Roberts:1994dr}, which are being used to develop a systematic, symmetry-preserving, continuum approach to the strong-interaction bound-state problem \cite{Chang:2011vu, Bashir:2012fs, Roberts:2015lja, Horn:2016rip, Eichmann:2016yit, Burkert:2017djoFBS}.

The Poincar\'e-covariant Faddeev equation for baryons is typically treated in a quark-diquark approximation, where the diquark correlations are nonpointlike and dynamical \cite{Segovia:2015ufa}.  This amounts to a simplified treatment of the scattering problem in the two-body subchannels (as explained, \emph{e.g}.\ in Ref.\,\cite{Hecht:2002ej}, Sec.\,II.A.2), which is founded on an observation that the same interaction which describes colour-singlet mesons also generates diquark correlations in the colour-antitriplet $(\bar 3)$ channel \cite{Cahill:1987qr, Maris:2002yu, Bi:2015ifa}.  Whilst the diquarks do not survive as asymptotic states, \emph{viz}.\ they are absent from the strong interaction spectrum \cite{Bender:1996bb, Bhagwat:2004hn}, the attraction between the quarks in the $\bar 3$ channel sustains a system in which two quarks are always correlated as a colour-$\bar 3$ pseudoparticle, and binding within the baryon is effected by the iterated exchange of roles between the bystander and diquark-participant quarks.

The quark-diquark approach to the spectrum and interactions of baryons continues to be applied broadly and recent applications include:
nucleon and $\Delta$-baryon elastic and transition form factors \cite{Segovia:2014aza, Roberts:2015dea};
the proton-to-Roper-resonance transition \cite{Segovia:2015hra, Roberts:2018hpf}, extending to a flavour separation of the associated form factors \cite{Segovia:2016zyc, Chen:2018nsgFBS};
%
%
structure studies of negative-parity baryon resonances \cite{Eichmann:2016jqx, Eichmann:2016nsu, Lu:2017cln, Chen:2017pse};
parton distribution amplitudes of the nucleon and Roper resonance \cite{Mezrag:2017znp};
and the spectrum and structure of octet and decuplet baryons and their positive-parity excitations \cite{Chen:2019fznFBS}.
Some of these studies are reviewed in this volume \cite{SegoviaFaddeev}.

Notably, the predictions of the quark-diquark Faddeev equation framework are consistent with experiments, including those associated with modern measurements of nucleon resonance electrocouplings \cite{Mokeev:2015lda, Mokeev:2018zxt}.  Moreover, the use of such methods, with largely unfettered application to a wide range of static and dynamic hadron properties, is of growing importance, considering recent advances in charting the spectrum of excited nucleons using electromagnetic probes.  For instance: a recent global multi-channel analysis of exclusive meson photoproduction revealed evidence for several new baryon states \cite{Anisovich:2017bsk}; and combined studies of charged double-pion photo- and electro-production data provide strong indications for another new baryon, \emph{i.e}.\ $N^\prime (1730) 3/2^+$ \cite{Ripani:2002ss, Mokeev:2015moa, Golovatch:2018hjk}.

The first treatment of the Poincar\'e-covariant Faddeev equation for the nucleon to eschew the quark-diquark approximation is described in Ref.\,\cite{Eichmann:2009qa}.  Regarding the nucleon mass, it revealed that the quark-diquark picture is accurate at the level of 5\%.  A variety of applications ensued: nucleon electromagnetic form factors
\cite{Eichmann:2011vu};  
masses of the $\Delta$- and $\Omega$-baryons
\cite{SanchisAlepuz:2011jn};  
nucleon axial and pseudoscalar form factors
\cite{Eichmann:2011pv};  
a spectrum of ground-state octet and decuplet baryons
\cite{Sanchis-Alepuz:2014sca}  
and their electromagnetic form factors
\cite{Sanchis-Alepuz:2015fcg};  
masses of low-lying nucleon excited states
\cite{Eichmann:2016hgl};  
and electromagnetic transition form factors between ground-state octet and decuplet baryons
\cite{Sanchis-Alepuz:2017mir}.  
Significant algebraic and computational effort was required to complete these studies; and the results are instructive and promising, indicating that the framework is potentially capable of drawing a traceable connection between QCD and the many baryon observables that are being made accessible by modern facilities.

This approach has also been used recently to provide Poincar\'e-covariant calculations of: the spectrum of $J^P=3/2^+$ baryons, including those with heavy-quarks, and their first positive-parity excitations \cite{Qin:2018dqp}; and the proton's tensor charges \cite{Wang:2018kto}.
Herein we describe an extension of the spectrum calculation to include all ground-state $J^P=1/2^+, 3/2^+$ baryons, including systems with one or more heavy-quarks, and their first positive-parity excitations and negative-parity parters.

Section~\ref{SecTBAE} reviews the Faddeev equation for baryons and introduces the leading-order truncation that enables symmetry-preserving solutions to be obtained.  The associated exchange-interaction kernel is described in Sec.\,\ref{SeccalG}, with some aspects of the dressed-quark propagators, used to complete the Faddeev kernel, detailed in Sec.\,\ref{GapEq}.  Our spectrum calculation and results are described and explained in Sec.\,\ref{baryonspectrum}.  Section~\ref{epilogue} provides a summary and indicates some new directions.

\section{Three-Body Amplitudes and Equations}
\label{SecTBAE}
We begin by sketching some features of the Poincar\'e-covariant Faddeev equation and its solution, using the isospin $I=1/2$, $J^P=1/2^+$ nucleon as an exemplar.  The Faddeev amplitude for this system can be written as follows:
\begin{align}
 \, _{c_1 c_2 c_3} \mathbf{\Psi}^{\alpha_1\alpha_2\alpha_3,\delta}_{\iota_1 \iota_2 \iota_3,\iota}(p_1,p_2,p_3;P)
& = \tfrac{1}{\surd 6} \varepsilon_{c_1 c_2 c_3}
{\Psi}^{\alpha_1\alpha_2\alpha_3,\delta}_{\iota_1 \iota_2 \iota_3,\iota}(p_1,p_2,p_3;P)  \,,
\label{FaddeevAmp}
\end{align}
where
$c_{1,2,3}$ are colour indices;
$\alpha_{1,2,3}$, $\delta$ are spinor indices for the three valence quarks and nucleon, respectively;
$\iota_{1,2,3}$, $\iota$ are analogous isospin indices;
and $P=p_1+p_2+p_3$, $P^2 = -M_{N}^2$, where $M_N$ is the nucleon mass and $p_{1,2,3}$ are the valence-quark momenta.  (Our Euclidean metric conventions are explained in Appendix\,B of Ref.\,\cite{Segovia:2014aza}.)

With colour factorised from the amplitude in Eq.\,\eqref{FaddeevAmp}, then
${\Psi}_{\iota_1 \iota_2 \iota_3,\iota}^{\alpha_1\alpha_2\alpha_3,\delta}(p_1,p_2,p_3;P)$ describes momentum-space$+$spin$+$isospin correlations in the nucleon and must be symmetric under the interchange of any two valence quarks, including cyclic permutations, \emph{e.g}.\
\begin{equation}
{\Psi}^{\alpha_1\alpha_2\alpha_3,\delta}_{\iota_1 \iota_2 \iota_3,\iota}(p_1,p_2,p_3;P) = {\Psi}^{\alpha_1\alpha_3\alpha_2,\delta}_{\iota_1 \iota_3 \iota_2,\iota}(p_1,p_3,p_2;P)\,.
\end{equation}

The structure of this matrix-valued function is nontrivial in a Poincar\'e-covariant treatment.  Considering isospin, there are three valence quarks in the fundamental representation of $SU(2)$:
\begin{equation}
\label{YoungD}
2\otimes 2\otimes 2 = 4 \oplus 2 \oplus 2\,.
\end{equation}
The fully-symmetric four-dimensional irreducible representation (irrep) is associated with the $J^P=3/2^+$ $\Delta$-baryon and will be used later.  In terms of valence-quark flavours, the two mixed-symmetry $I=1/2$ two-dimensional irreps can be depicted thus:
\begin{equation}
\begin{array}{c|c|c}
       & I_z =\frac{1}{2} &  I_z = -\frac{1}{2} \\ \hline
 \mathsf{F}_0      & \tfrac{1}{\surd 2} (udu - duu) & \tfrac{1}{\surd 2} (udd - dud) \\
 \mathsf{F}_1      &  -\tfrac{1}{\surd 6}(udu+duu-2uud) & \tfrac{1}{\surd 6}(udd+dud-2ddu)
\end{array}\,.
\end{equation}
Defining a quark isospin vector $\mathsf{f}=(u,d)$, then this array can be expressed compactly via matrices:
\begin{equation}
\label{DiquarkSeeds}
D_0 = \tfrac{i}{\surd 2} \tau^2 \otimes \tau^0\,,\;
D_1 = -\tfrac{i}{\surd 6} \tau^i \tau^2 \otimes \tau^i\,,
\end{equation}
with $\tau^0 = {\rm diag}[1,1]$ and $\{\tau^i,i=1,2,3\}$ being Pauli matrices, \emph{e.g}.\ the bottom-left entry is
\begin{equation}
(\mathsf{f}\, \mathsf{f}^{\rm T}) D_1  (\mathsf{f}\, \mathsf{p}^{\rm T}) = -\tfrac{i}{\surd 6} \mathsf{f} \tau^i \tau^2 \mathsf{f}^{\rm T} \, \mathsf{f} \tau^i \mathsf{p}^{\rm T},
\end{equation}
where $\mathsf{p}=(1,0)$ represents the $I_z=+1/2$ proton and $(\cdot)^{\rm T}$ indicates matrix-transpose.  Notably, with respect to the first two labels, $D_0$ relates to isospin-zero and $D_1$ to isospin-one; and differences between quark-quark scattering in these channels can provide the seed for formation of diquark correlations within baryons \cite{Segovia:2015ufa}.  Such differences do exist, \emph{e.g}.\ only $u$-$d$ scattering possesses an attractive isospin-zero channel.  (It is anticipated that continuing examination of the Faddeev equation's solutions and their dependence on the structure of the kernel will deliver an understanding of the dynamics behind the emergence of diquark correlations within baryons.  Such efforts will likely benefit from the use of high-performance computing.)

Labelling the valence quarks by $\{i,j,k\}$, each taking a distinct value in $\{1,2,3\}$, then under $i \leftrightarrow j$
\begin{equation}
\left[\begin{array}{c}
\mathsf{F}_0 \\ \mathsf{F}_1 \end{array} \right]
\to
\left[\begin{array}{c}
\mathsf{F}_0^\prime \\ \mathsf{F}_1^\prime \end{array} \right]
=
{\mathpzc E}_k
\left[\begin{array}{c}
\mathsf{F}_0 \\ \mathsf{F}_1 \end{array} \right]\,,
\end{equation}
where ${\mathpzc E}_k$ is the associated exchange operator.  In general, owing to the mixed symmetry of these irreps, $\mathsf{F}_{0,1}^\prime \neq \mathsf{F}_{0,1}$.  Define in addition, therefore, a momentum-space$+$spinor doublet with the following transformation properties:
\begin{equation}
\left[ \Psi_0 \Psi_1 \right] \to \left[ \Psi_0 \Psi_1 \right] {\mathpzc E}_k^{\rm T}.
\end{equation}
Consequently, the momentum-space$+$spinor$+$isospin combination
\begin{align}
\Psi(p_1,p_2,p_3;P)
&= \Psi_0 (p_1,p_2,p_3;P) \mathsf{F}_0 + \Psi_1(p_1,p_2,p_3;P) \mathsf{F}_1
\label{FinalAmplitude}
\end{align}
is invariant under the exchange of any two quark labels.\footnote{Given this ``doublet'' structure, $64+64=128$ independent scalar functions are required to completely describe a nucleon Faddeev amplitude: see Appendix\,B in Ref.\,\cite{Eichmann:2011vu} for more details.}  This feature is a statement of the fact that a Poincar\'e-covariant treatment of the nucleon does not typically admit a solution in which the momentum-space behaviour is independent of the spin-isospin structure; or, equivalently, that using a Poincar\'e-covariant framework, the $d$-quark contribution to a nucleon's form factor or kindred property is not simply proportional to the $u$-quark contribution.

\begin{figure}[!t]
\begin{center}
\includegraphics[clip,width=0.66\linewidth]{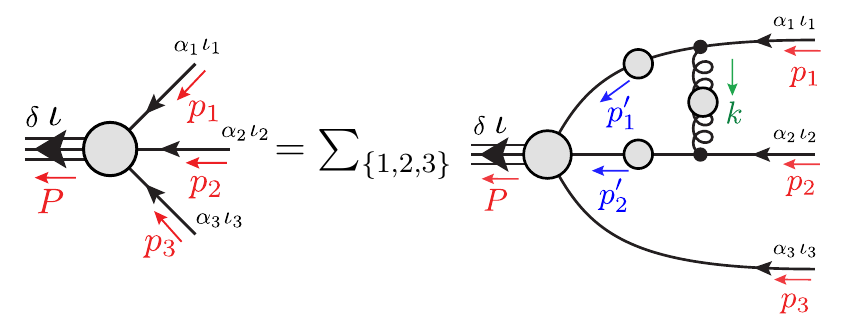}
\end{center}
\vspace*{-2em}
\caption{\label{FEimage}
Three-body equation in Eq.\,\eqref{eq:faddeev0}, used herein to compute baryon masses and bound-state amplitudes.
Amplitude: vertex on the left-hand-side;
spring with shaded circle: quark-quark interaction kernel in Eq.\,\eqref{KDinteraction};
and solid line with shaded circle: dressed-propagators for scattering quarks, obtained by solving a gap equation with the same interaction (Sec.\,\ref{GapEq}).
}
\end{figure}

As indicated in the Introduction, the continuum bound-state problem is naturally embedded in the DSE approach to strong-QCD.  A tractable system of DSEs is only obtained once a truncation scheme is specified; and a systematic, symmetry-preserving approach is described in Refs.\,\cite{Munczek:1994zz, Bender:1996bb, Binosi:2016rxz}.  The leading-order term is the rainbow-ladder (RL) truncation.  It is known to be accurate for ground-state light-quark vector- and isospin-nonzero-pseudoscalar-mesons, and related ground-state octet and decuplet baryons \cite{Chang:2011vu, Bashir:2012fs, Roberts:2015lja, Horn:2016rip, Eichmann:2016yit} because corrections largely cancel in these channels owing to the preservation of relevant Ward-Green-Takahashi identities \cite{Ward:1950xp, Green:1953te, Takahashi:1957xn} ensured by the scheme \cite{Munczek:1994zz, Bender:1996bb, Binosi:2016rxz}.
To obtain the nucleon amplitude in Eq.\,\eqref{FinalAmplitude}, we therefore consider the following RL-truncation three-body equation, depicted in Fig.\,\ref{FEimage}:
\begin{subequations}
 \label{eq:faddeev0}
\begin{align}
{\Psi}_{\iota_1 \iota_2 \iota_3,\iota}^{\alpha_1\alpha_2\alpha_3,\delta}&(p_1,p_2,p_3)
= \sum_{j=1,2,3} \big[ {\mathscr K} S S \Psi \big]_j\,,
\label{eq:faddeev01}\\
\big[ {\mathscr K} S S \Psi \big]_3 & =
\int_{dk} \mathscr{K}_{\iota_1 \iota_1^\prime \iota_2 \iota_2^\prime}^{\alpha_1\alpha_1',\alpha_2\alpha_2'}(p_1,p_2;p_1',p_2') 
%
S_{\iota_1^\prime \iota_1^{\prime\prime}}^{\alpha_1'\alpha_2''}(p_1')
S_{\iota_2^\prime \iota_2^{\prime\prime}}^{\alpha_2'\alpha_2''}(p_2')
{\Psi}_{\iota_1'' \iota_2'' \iota_3;\iota}^{\alpha_1''\alpha_2''\alpha_3;\delta}(p_1',p_2',p_3) \,,
\end{align}
\end{subequations}
where $\int_{dk}$ represents a translationally-invariant definition of the four-dimensional integral and $\big[ {\mathscr K} S S \Psi\big]_{1,2}$ are obtained from $\big[ {\mathscr K} S S \Psi \big]_3$ by cyclic permutation of indices. 

\section{Two-Body Interaction}
\label{SeccalG}
The key element in analyses of the continuum bound-state problem for hadrons is the quark-quark scattering kernel.  In RL truncation that can be written ($k = p_1-p_1^\prime = p_2^\prime -p_2$):
\begin{equation}
\label{KDinteraction}
\mathscr{K}_{\alpha_1\alpha_1',\alpha_2\alpha_2'}  = {\mathpzc G}_{\mu\nu}(k) [i\gamma_\mu]_{\alpha_1\alpha_1'} [i\gamma_\nu]_{\alpha_2\alpha_2'}\,,\quad
 {\mathpzc G}_{\mu\nu}(k)  = \tilde{\mathpzc G}(k^2) T_{\mu\nu}(k)\,,
\end{equation}
where $k^2 T_{\mu\nu}(k) = k^2 \delta_{\mu\nu} - k_\mu k_\nu$.
Thus, in order to define all elements in Eq.\,\eqref{eq:faddeev0} and hence the bound-state problem, it remains only to specify $\tilde{\mathpzc G}$.  Two decades of study have yielded the following form \cite{Qin:2011dd, Qin:2011xq} ($s=k^2$):
\begin{align}
\label{defcalG}
 \tfrac{1}{Z_2^2}\tilde{\mathpzc G}(s) & =
 \frac{8\pi^2}{\omega^4} D e^{-s/\omega^2} + \frac{8\pi^2 \gamma_m \mathcal{F}(s)}{\ln\big[ \tau+(1+s/\Lambda_{\rm QCD}^2)^2 \big]}\,,
\end{align}
where \cite{Qin:2018dqp}: $\gamma_m=12/(33-2N_f)$, $N_f=5$; $\Lambda_{\rm QCD}=0.36\,$GeV; $\tau={\rm e}^2-1$; and $s{\cal F}(s) = \{1 - \exp(-s/[4 m_t^2])\}$, $m_t=0.5\,$GeV.  $Z_2$ is the dressed-quark wave function renormalisation constant.\footnote{In all calculations herein, we employ a mass-independent momentum-subtraction renormalisation scheme for all relevant DSEs, implemented by making use of the scalar Ward-Green-Takahashi identity and fixing all renormalisation constants in the chiral limit \cite{Chang:2008ec}, with renormalisation scale $\zeta=19\,$GeV$=:\zeta_{19}$.}  
The development of Eqs.\,\eqref{KDinteraction}, \eqref{defcalG} is summarised in Ref.\,\cite{Qin:2011dd} and their connection with QCD is described in Ref.\,\cite{Binosi:2014aea}.

Computations \cite{Qin:2011dd, Qin:2011xq, Chen:2018rwz} reveal that observable properties of light-quark ground-state vector- and flavour-nonsinglet pseudoscalar-mesons are practically insensitive to variations of $\omega \in [0.4,0.6]\,$GeV, so long as
\begin{equation}
 \varsigma^3 := D\omega = {\rm constant}.
\label{Dwconstant}
\end{equation}
This feature extends to numerous properties of the nucleon and $\Delta$-baryon \cite{Eichmann:2008ef, Eichmann:2012zz, Wang:2018kto}.  The value of $\varsigma$ is typically chosen so as to reproduce the measured value of the pion's leptonic decay constant, $f_\pi$; and in RL truncation this requires
\begin{equation}
\label{varsigmalight}
\varsigma  =0.80\,{\rm GeV.}
\end{equation}
We will subsequently employ $\omega=0.5\,$GeV, the midpoint of the domain of insensitivity.

It is also worth looking at Eq.\,\eqref{defcalG} from a different perspective \cite{Binosi:2014aea, Binosi:2016nme, Rodriguez-Quintero:2018wma}.  Namely, one can sketch a connection with QCD's renormalisation-group-invariant process-independent effective-charge by writing
\begin{equation}
\label{alphaRL}
\tfrac{1}{4\pi}\tilde{\mathpzc G}(s) \approx \frac{\tilde\alpha_{\rm PI}(s)}{s + \tilde m_g^2(s)}\,,\;
 m_g^2(s) = \frac{\tilde m_0^4}{s + \tilde m_0^2}\,,
\end{equation}
and extract $\tilde\alpha_0:=\tilde\alpha_{\rm PI}(0)$, $\tilde m_0$ via a least-squares fit on an infrared domain: $s\lesssim M_N^2$.  This yields
\begin{equation}
\label{RLcouplings}
\tfrac{1}{\pi}\tilde\alpha_0^{\rm RL} = 9.7\,,\; \tilde m_0^{\rm RL} = 0.54\,{\rm GeV}\,,\;
\end{equation}
$\alpha_0^{\rm RL}/\pi/[m_0^{\rm RL}]^2 \approx 33\,$GeV$^{-2}$.  Comparison of these values with those predicted via a combination of continuum and lattice analyses of QCD's gauge sector \cite{Binosi:2016nme, Rodriguez-Quintero:2018wma}: $\alpha_0/\pi \approx 1.0$, $m_0 \approx 0.5\,$GeV, $\alpha_0/\pi/m_0^2 \approx 4.2\,$GeV$^{-2}$, confirms an earlier observation \cite{Binosi:2014aea} that the RL interaction defined by Eqs.\,\eqref{KDinteraction}, \eqref{defcalG} has the right shape, but is an order-of-magnitude too large in the infrared.  As explained elsewhere \cite{Chang:2009zb, Chang:2010hb, Chang:2011ei}, this is because Eq.\,\eqref{KDinteraction} suppresses all effects associated with dynamical chiral symmetry breaking (DCSB) in bound-state equations \emph{except} those expressed in $\tilde{\mathpzc G}(k^2)$, and therefore a description of hadronic phenomena can only be achieved by overmagnifying the gauge-sector interaction strength at infrared momenta.

It should also be noted that in choosing the scale in Eq.\,\eqref{varsigmalight} so as to describe a given set of light-hadron observables in RL truncation, some effects of resonant (meson cloud) contributions to light-hadron static properties are implicitly included \cite{Eichmann:2008ae}.  We capitalise on this feature herein; and return to this point below.

We subsequently also consider systems involving heavy-quarks, so it is pertinent to remark that RL truncation has also been explored in connection with heavy-light mesons and heavy-quarkonia \cite{Bhagwat:2004hn, Hilger:2014nma, Ding:2015rkn, Gomez-Rocha:2016cji, Chen:2016bpj, Hilger:2017jti, Binosi:2018rht}.  Those studies reveal that improvements to RL are critical in heavy-light systems; and a RL-kernel interaction strength fitted to pion properties alone is not optimal in the treatment of heavy quarkonia.  Both observations are readily understood, but we focus on the latter because it is most relevant herein.

Recall, therefore, that for meson bound-states it is now possible \cite{Chang:2009zb, Chang:2010hb, Chang:2011ei} to employ sophisticated kernels which overcome many of the weaknesses of RL truncation.  The new technique is symmetry preserving and has an additional strength, \emph{i.e}.\ the capacity to express DCSB nonperturbatively in the integral equations connected with bound-states.  Owing to this feature, the scheme is described as the ``DCSB-improved'' or ``DB'' truncation.  In a realistic DB truncation, $\varsigma^{\rm DB} \approx 0.6\,$GeV; a value which coincides with that predicted by solutions of QCD's gauge-sector gap equations \cite{Binosi:2014aea, Binosi:2016wcx, Binosi:2016nme, Rodriguez-Quintero:2018wma}.
Straightforward analysis shows that corrections to RL truncation largely vanish in the heavy+heavy-quark limit;
%
%
hence the aforementioned agreement entails that RL truncation should provide a sound approximation for systems involving only heavy-quarks so long as one employs $\varsigma^{\rm DB}$ as the infrared mass-scale.  In heavy-quark systems we therefore employ Eqs.\,\eqref{KDinteraction}, \eqref{defcalG} as obtained using
\begin{equation}
\label{varsigmaQ}
\varsigma_Q = 0.6\,{\rm GeV}\,.
\end{equation}

\section{Dressed-quark Propagator}
\label{GapEq}
The kernel of the Faddeev equation, Eq.\,\eqref{eq:faddeev0}, is complete once the dressed-quark propagator is known; and to ensure a symmetry-preserving analysis in the present case, this should be computed from the following (rainbow truncation) gap equation ($q\in \{u,d,s,c,b\}$ labels the quark flavour):
\begin{subequations}
\label{gendseN}
\begin{align}
S_q^{-1}(k) & = i\gamma\cdot k \, A_q(k^2) + B_q(k^2)
= [i\gamma\cdot k + M_q(k^2)]/Z_q(k^2)\,, \\
& = Z_2 \,(i\gamma\cdot k + m_q^{\rm bm}) +  \int_{d\ell}\!\!
 {\mathpzc G}_{\mu\nu}(k-\ell)\frac{\lambda^a}{2}\gamma_\mu S_q(\ell) \frac{\lambda^a}{2} \gamma_\nu \,,
\end{align}
\end{subequations}
using the interaction specified in connection with Eqs.\,\eqref{KDinteraction}, \eqref{defcalG}.  Following Ref.\,\cite{Maris:1997tm}, this gap equation is now readily solved, and we adapt algorithms from Ref.\,\cite{Krassnigg:2009gd} when necessary.

\begin{table}[t]
\caption{\label{obsuds}
Computed values for a range of light-quark-hadron properties (masses and leptonic decay constants), obtained using the quark-quark scattering kernel described in Sec.\,\ref{SeccalG} to specify the relevant gap- and Bethe-Salpeter-equations.  The interaction scale is stated in Eq.\,\eqref{varsigmalight}; and the current-quark masses in Eq.\,\eqref{currentuds} were chosen to reproduce the empirical values of $m_\pi=0.14\,$GeV, $m_K=0.50\,$GeV.  ($Z_2(\zeta_{19})\approx 1$.)
(Computed results drawn from Ref.\,\cite{Qin:2018dqp} and experimental values drawn from Ref.\,\cite{Tanabashi:2018oca}.  All quantities listed in GeV.)}
\begin{center}
\begin{tabular*}
{\hsize}
{|l@{\extracolsep{0ptplus1fil}}
|l@{\extracolsep{0ptplus1fil}}
l@{\extracolsep{0ptplus1fil}}
l@{\extracolsep{0ptplus1fil}}
l@{\extracolsep{0ptplus1fil}}
l@{\extracolsep{0ptplus1fil}}
l@{\extracolsep{0ptplus1fil}}
l@{\extracolsep{0ptplus1fil}}
l|@{\extracolsep{0ptplus1fil}}}\hline
       & $f_\pi$ & $f_K$ & $m_\rho$ & $f_\rho$ & $m_{K^\ast}$ & $f_{K^\ast}$ & $m_\phi$ & $f_\phi$   \\\hline
herein\; & $0.094$ & $0.11$ & $0.75$ & $0.15$ & $0.95$ & $0.18$ & $1.09$ & $0.19$ \\
expt. & $0.092$ & $0.11$ & $0.78$ & $0.15$ & $0.89$  & $0.16$ & $1.02$ & $0.17$ \\\hline
\end{tabular*}
\end{center}
\end{table}

All that remains to be specified are the Higgs-generated current-quark masses, $m_q$.   We work in the isospin symmetric limit, with $m_l:=m_u=m_d$, and find that the choices
\begin{equation}
\label{currentuds}
m_{l}^{\zeta_{19}} = 3.3\,{\rm MeV}\,,\; m_s^{\zeta_{19}} = 74.6\,{\rm MeV}\,,
\end{equation}
when used to determine the gap equation solutions that feed into the Bethe-Salpeter equations, yield the results in Table~\ref{obsuds}.\footnote{We reiterate that the mass-scale in Eq.\,\eqref{varsigmalight} makes no allowance for the effect of corrections to RL truncation on light-hadron observables.  This issue is canvassed elsewhere \cite{Eichmann:2008ae}, with the following conclusion: for systems in which orbital angular momentum does not play a big role, the impact of such corrections may largely be absorbed in a redefinition of this scale.  With some revisions, we adapt this idea below to systems with angular momentum and to radial excitations.
}
The values in Eq.\,\eqref{currentuds} correspond to renormalisation-group-invariant masses $\hat m_{u,d}=6.3\,$MeV, $\hat m_s=146\,$MeV; one-loop-evolved masses at 2\,GeV of
\begin{equation}
m_{l}^{2\,{\rm GeV}}=4.8\,{\rm MeV}\,, \; m_{s}^{2\,{\rm GeV}}=110\,{\rm MeV}\,;
\end{equation}
Euclidean constituent quark masses
\begin{equation}
M^E_{l}=0.41\,{\rm GeV},\quad M^E_s=0.57\,{\rm GeV},
\end{equation}
defined via $M_q^E = \{k|M_q(k)=k\}$, where $M_q(k)$ is the nonperturbative solution of the appropriate gap equation;
and give $\hat m_s/\hat m_{u=d}=23$.
Evidently, our current-quark masses are compatible with modern estimates by other means \cite{Tanabashi:2018oca}.

Bound states involving heavy quarks, $Q=c,b$, were analysed in Ref.\,\cite{Qin:2018dqp} using RL truncation with $\varsigma_Q$ in Eq.\,\eqref{varsigmaQ} and $\omega_Q = 0.8\,$GeV, as appropriate for the heavy-quark sector \cite{Chen:2016bpj, Binosi:2018rht}, and current-quark masses
\begin{equation}
\label{currentQR} 
m_{c}^{\zeta_{19}} = 0.83\,{\rm GeV}\,,\; m_b^{\zeta_{19}} = 3.66\,{\rm GeV}\,.
\end{equation}
These masses correspond
to renormalisation-group-invariant masses $\hat m_{c}=1.64\,$GeV, $\hat m_b=7.30\,$GeV; one-loop-evolved masses at 2\,GeV of
\begin{equation}
\label{mQOO}
m_{c}^{2\,{\rm GeV}}=1.24\,{\rm GeV}\,, \; m_{b}^{2\,{\rm GeV}}=5.52\,{\rm GeV};
\end{equation}
Euclidean constituent quark masses
\begin{equation}
M^E_c=1.35\,{\rm GeV}, \quad M^E_b=4.28\,{\rm GeV};
\end{equation}
and give $\hat m_c/\hat m_s=11$, $\hat m_b/\hat m_s=50$.
As with the $u,d,s$ masses, these values are compatible with other contemporary estimates \cite{Tanabashi:2018oca}.

\section{Baryon Spectrum}
\label{baryonspectrum}
\subsection{$J^P = 1/2^+$, $3/2^+$ Ground States}
Eq.\,\eqref{eq:faddeev0} can now be solved for the nucleon mass and Poincar\'e-covariant bound-state amplitude,\footnote{The formulation of this problem and efficient solution methods are detailed, \emph{e.g}.\  in Ref.\,\cite{Eichmann:2011vu}, Appendices A--C, and Ref.\,\cite{Qin:2018dqp}, Appendices A, B.}
using the interaction described in Sec.\,\ref{SeccalG} and the dressed-quark propagator from Sec.\,\ref{GapEq}.  We obtain
\begin{equation}
\label{MassN}
m_{N} = 0.948^{(05)}_{(11)}\,{\rm GeV}\,,
\end{equation}
where the fluctuation ($\lesssim 1$\%) indicates the sensitivity of this prediction to the variation $\omega = 0.5 \mp 0.05$.  Importantly, no parameters were varied to obtain the value in Eq.\,\eqref{MassN}: it follows once the scale in Eq.\,\eqref{Dwconstant} is chosen and $m_l$ is fixed to give the empirical pion mass, Eq.\,\eqref{currentuds}.

The momentum-space$+$spin$+$flavour structures of the remaining members of the ground-state baryon octet are readily obtained by generalising the discussion associated with Eqs.\,\eqref{YoungD}\,--\,\eqref{FinalAmplitude} to flavour-$SU(3)$.  The $\Sigma^+=(uus)_{I=1}$, $\Sigma^-=(dds)_{I=1}$, $\Xi^0=(uss)_{I=1/2}$, $\Xi^- = (dss)_{I=1/2}$ are particularly simple because, \emph{e.g}.\ the $\Sigma^+$ is simply obtained from the proton by swapping $d\to s$.   In the isospin-symmetry limit, $\Sigma^0=(uds)_{I=1}$ introduces no complications because it is degenerate with $\Sigma^{\pm}$.
The $\Lambda^0=(uds)_{I=0}$, on the other hand, is different because the wave function must express $I=0$; and, moreover, it is empirically 6\% lighter than the $\Sigma^0$.
This splitting is readily explained by the quark-diquark picture of baryons \cite{Lu:2017cln, Chen:2019fznFBS}, in which the $\Lambda_{I=0}^0$ contains more of the lighter $J^P=0^+$ diquark correlations than the $\Sigma_{I=1}^0$, \emph{viz}.\ the wave functions of these two systems exhibit dynamically-generated differences.
However, using Eq.\,\eqref{eq:faddeev0}, \emph{i.e}.\ in a pure RL truncation of the three-body problem, $\Lambda_{I=0}^0$ and $\Sigma_{I=1}^0$ are degenerate.  (Isospin partners in the meson spectrum are also degenerate in RL truncation \cite{Roberts:1988yz, Hollenberg:1992nj, Pichowsky:1999mu}.)
%

We have not yet developed our methods to the point where we can directly solve a Faddeev equation for the mixed-flavour $\Sigma$ and $\Xi$ baryons.  However, so far as computing the spectrum in RL truncation is concerned, that may not be necessary.  Instead, if accuracy at a level of $\lesssim 5$\% is sufficient, then one can employ the equal-spacing scheme introduced in Ref.\,\cite{Qin:2018dqp}, which we now describe.

\begin{figure}[!t]
\begin{center}
\includegraphics[clip,width=0.66\linewidth]{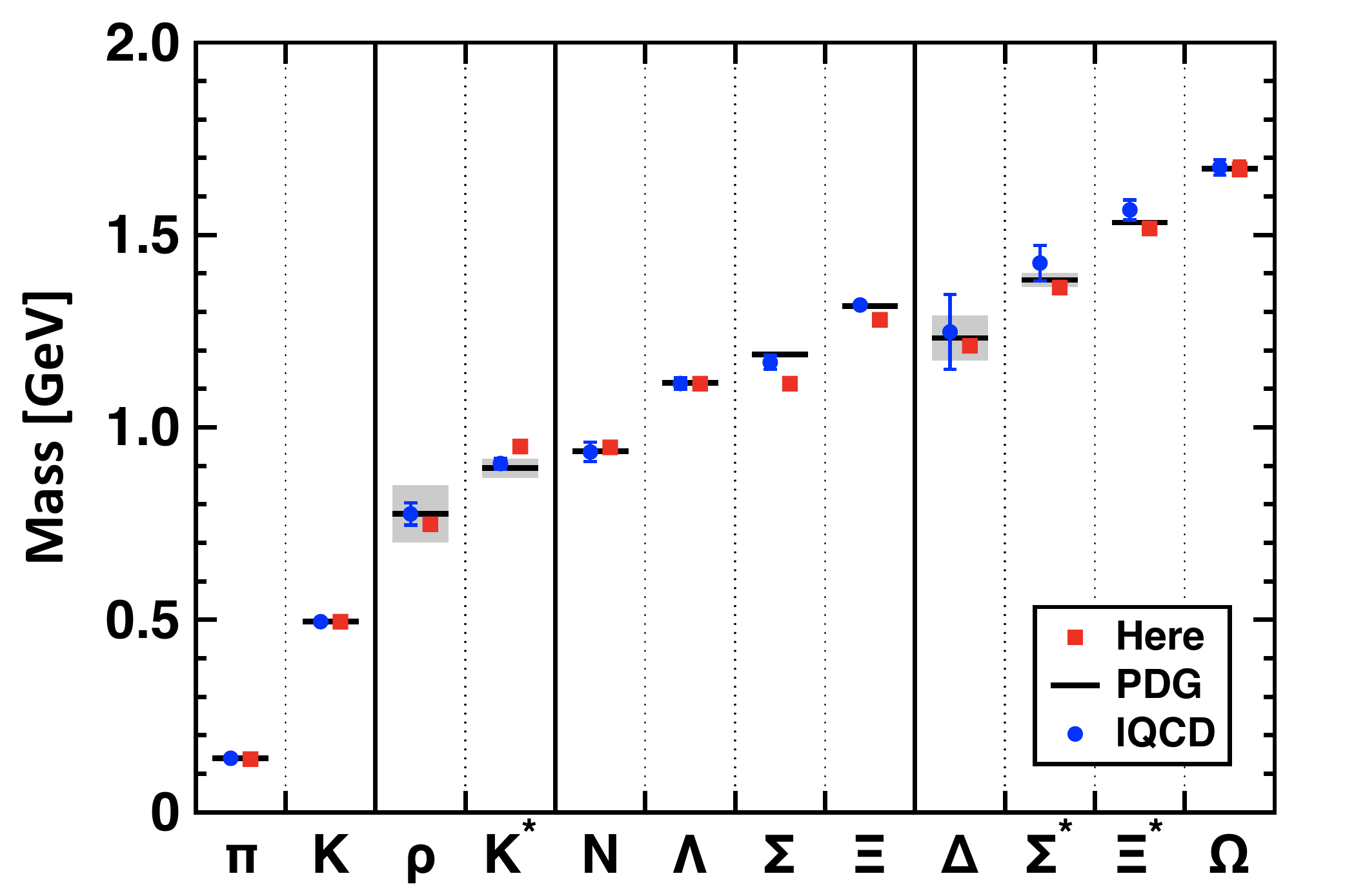}
\end{center}
\caption{\label{810MassComparison}
Masses of pseudoscalar and vector mesons, and ground-state positive-parity octet and decuplet baryons calculated herein (squares, red) compared with: experiment (black bars, with decay-widths of unstable states shaded in grey); and masses computed using lQCD \cite{Durr:2008zz}.
The sensitivity of our results to a 10\% variation in our single parameter ($\omega$ in Eq.\,\eqref{varsigmalight}, with $D\omega=\,$constant) is smaller than the symbol.  The light- and strange-quark current masses, Eq.\,\eqref{currentuds}, were fitted to the pion and kaon masses.  All other results are predictions.
}
\end{figure}

Suppose one has three distinct flavours of degenerate quarks.  In the isospin-symmetry limit, the nucleon described above is such a system; and we have computed its wave function and mass, Eq.\,\eqref{MassN}.  The same Faddeev equation codes can be used to compute the mass of this bound-state when all three quarks possess the $s$-quark mass, with the result:
\begin{equation}
m_{N_s} ({\rm GeV}) = 1.444^{(08)}_{(17)}\,.
\end{equation}
Now in the spirit of the equal spacing rule (ESR) \cite{Okubo:1961jc, GellMann:1962xb}, which RL truncation results for hadron masses and decay constants follow to a good approximation, we define the following $u$- and $s$-quark spectrum-constituent masses for the ground-state octet:
\begin{equation}
\label{OctetMq}
M_u^{\underline{8} 0}  = \tfrac{1}{3} m_N  = 0.316^{(2)}_{(4)}\,,\;
M_s^{\underline{8} 0}  = \tfrac{1}{3} m_{N_s} = 0.482^{(3)}_{(6)} \,,
\end{equation}
and estimate
\begin{equation}
m_\Lambda = m_\Sigma = 2 M_u^{\underline{8} 0} + M_s^{\underline{8} 0} = 1.114^{(06)}_{(13)}\,,\;
m_\Xi = M_u^{\underline{8} 0} + 2 M_s^{\underline{8} 0} = 1.279^{(07)}_{(15)}\,.
\end{equation}
These predictions are compared in Fig.\,\ref{810MassComparison} with experiment \cite{Tanabashi:2018oca} and results from numerical simulations of lattice-regularised QCD (lQCD) \cite{Durr:2008zz}.

The analogue of Eq.\,\eqref{eq:faddeev0} appropriate to decuplet baryons is described in Ref.\,\cite{Qin:2018dqp}; and the procedure just described can also be applied to this case.  By direct computation, with the same interaction parameter and current-quark masses specified above, one finds\footnote{As it was above, in all subsequent cases the sensitivity to $\pm 10$\% variations of $\omega$ in Eq.\,\eqref{varsigmalight}, with $D\omega=\,$constant, is uniformly $\lesssim 1$\%.  We therefore omit further mention of it hereafter.}
\begin{equation}
m_\Delta = 1.210\,,\;
m_\Omega = 1.670\,.
\end{equation}
Then, defining the following spectrum-constituent masses for the ground-state decuplet:
\begin{equation}
\label{DecupletMq}
M_u^{\underline{10} 0}  = \tfrac{1}{3} m_\Delta  = 0.403\,,\;
M_s^{\underline{10} 0}  = \tfrac{1}{3} m_{\Omega} = 0.557 \,,
\end{equation}
we compute
\begin{equation}
m_{\Sigma^\ast} = 2 M_u^{\underline{10} 0} + M_s^{\underline{10} 0} = 1.136\,,\;
m_{\Xi^\ast} = M_u^{\underline{10} 0} + 2 M_s^{\underline{10} 0} = 1.517\,.
\end{equation}
Our predictions for the decuplet masses are also depicted in Fig.\,\ref{810MassComparison} along with experimental values \cite{Tanabashi:2018oca} and results from lQCD \cite{Durr:2008zz}.

\begin{table}[t]
\caption{\label{SpectrumMq}
Channel-specific spectrum-constituent-quark masses computed herein and used to determine the masses of unlike-flavour baryons via equal-spacing rules.
The channels are specified thus: $(n,J^P)$, where $n=0$ indicates channel ground-state and $n=1$ is the first like-parity excited state.
(All masses listed in GeV.)}
\begin{center}
\begin{tabular*}
{\hsize}
{
|l@{\extracolsep{0ptplus1fil}}
|l@{\extracolsep{0ptplus1fil}}
l@{\extracolsep{0ptplus1fil}}
l@{\extracolsep{0ptplus1fil}}
l@{\extracolsep{0ptplus1fil}}
l|@{\extracolsep{0ptplus1fil}}
l@{\extracolsep{0ptplus1fil}}
l@{\extracolsep{0ptplus1fil}}
l@{\extracolsep{0ptplus1fil}}
l@{\extracolsep{0ptplus1fil}}
l|@{\extracolsep{0ptplus1fil}}
}
\hline
& $(0,\tfrac{1}{2}^+)\ $ & $(1,\tfrac{1}{2}^+)\ $ & $(0,\tfrac{1}{2}^-)\ $  & $(1,\tfrac{1}{2}^+)^\ast\ $       & $(0,\tfrac{1}{2}^-)^\ast\ $
& $(0,\tfrac{3}{2}^+)\ $ & $(1,\tfrac{3}{2}^+)\ $ & $(0,\tfrac{1}{2}^-)\ $  & $(1,\tfrac{3}{2}^+)^\ast\ $       & $(0,\tfrac{3}{2}^-)^\ast\ $\\\hline
$l\ $ & 0.316 & 0.426 & 0.381 & 0.480 & 0.514 & 0.403 & 0.487 & 0.466 & 0.542 & 0.575 \\
$s\ $ & 0.482 & 0.622 & 0.553 & 0.623 & 0.553 & 0.556 & 0.653 & 0.634 & 0.653 & 0.634 \\
$c\ $ & 1.552 & 1.690 & 1.638 & 1.690 & 1.638 & 1.587 & 1.717 & 1.676 & 1.717 & 1.676 \\
$b\ $ & 4.762 & 4.956 & 4.888 & 4.956 & 4.888 & 4.790 & 4.993 & 4.924 & 4.993 & 4.924 \\\hline
\end{tabular*}
\end{center}
\end{table}

A comparison of the values in Eq.\,\eqref{OctetMq} and Eq.\,\eqref{DecupletMq} indicates that these calculated spectrum-constituent masses depend on the channel, \emph{i.e}.\ the $J^P$ quantum numbers of the systems involved, and this difference diminishes with increasing $m_q$.  The same patterns are observed elsewhere \cite{Qin:2018dqp, Chen:2019fznFBS}, which also reveal that the ESR provides a good description of the masses of the first positive-parity excitations of these states, so long as the spectrum-constituent masses are recomputed accordingly.  These observations suggest that each like-$J^P$ excitation-level deriving from a flavour-$SU(N_f)$ baryon multiplet can be characterised by a set of $N_f$ level-specific spectrum-constituent-quark mass-scales; and the mass of each baryon in that level is well approximated by the sum of mass-scales dictated by the given baryon's valence-quark content.  As we shall see below, this generalised ESR provides a reliable means of computing the mass of a baryon constituted from two or more non-degenerate flavours given those of the degenerate-flavour states, whether they are known experimentally or theoretically.  All spectrum-constituent masses computed herein are listed in Table~\ref{SpectrumMq}.

The pattern revealed in Fig.\,\ref{810MassComparison} is interesting.  Seemingly, in order to explain the spectra of ground-state flavour-nonsinglet mesons and ground-state octet and decuplet baryons, at a level of 2.5\% mean-absolute-relative-difference, it is sufficient to employ a single mass-scale in the RL kernel that is fixed to reproduce the empirical value of the pion's leptonic decay constant.  This being so then, ignoring small isospin-breaking effects, these bound-states can all be understood as being composed of dynamically-dressed-quarks
\cite{Lane:1974he, Politzer:1976tv, Bhagwat:2003vw, Bowman:2005vx, Bhagwat:2006tu, Binosi:2016wcx} bound by the iterated exchange of gluons, themselves dressed and hence characterised by a running mass-scale that is large at infrared momenta \cite{Aguilar:2015bud}.\footnote{Notably, the mass of any given hadron is an integrated (long-wavelength) quantity; hence, not very sensitive to details of the system's wave function.  This feature plays a big role in the success of the ESR: so long as the centre-of-mass for each excitation-level is correctly set by the symmetry-preserving treatment of a broadly-sensible interaction, then a fair description of the spectrum should follow.  Dynamical quantities that evolve with a probe's momentum scale, \emph{e.g}.\ elastic and transition form factors, are needed to expose a bound-state's internal structure and so reveal details of the interaction which forms the composite system.}

Naturally, RL truncation is \emph{not} the complete picture: it is only the leading-order term in the systematic DSE truncation scheme described in Refs.\,\cite{Munczek:1994zz, Bender:1996bb, Binosi:2016rxz}.  As already noted, it \emph{works} for the ground states considered above because corrections largely cancel in these channels owing to the preservation of relevant Ward-Green-Takahashi identities and, hence, their effects can generally be absorbed in rescaling the interaction mass-scale \cite{Eichmann:2008ae}.  Consequently, as explained in connection with Eqs.\,\eqref{alphaRL}, \eqref{RLcouplings}, the interaction thus obtained, Eq.\,\eqref{defcalG} with Eq.\,\eqref{varsigmalight}, is only qualitatively consistent with QCD.  Nevertheless, Fig.\,\ref{810MassComparison} shows that the judicious use of RL truncation typically yields sound insights regarding ground-state hadrons.


\begin{table}[!t]
\caption{\label{MassesTableOneHalf}
Computed masses of $J=1/2$ baryons compared with experimental values \cite{Tanabashi:2018oca}, where known.  (Calculations assume isospin symmetry.)
States are labelled with the quark model name, drawn from Ref.\,\cite{Tanabashi:2018oca}, valence-quark content, and Faddeev equation identifications: $(n,1/2^P)$, where $n=0$ indicates channel ground-state and $n=1$ is the first like-parity excited state.  The numerical subscript indicates the table column number.
Columns labelled with an asterisk were computed as described in connection with Eqs.\,\eqref{sigmachangePP}, \eqref{sigmachangeRE}.
The mean-absolute-relative-difference between our best predictions (columns 6, 9, 10) and known experimental values is $3.6(2.7)$\%.
%
}
\begin{center}
\begin{tabular}{|c|c|l|l|l|r|r|r|r|r|}\hline
 \multicolumn{2}{|l}{\,} & \multicolumn{3}{|c}{ Empirical \cite{Tanabashi:2018oca} } & \multicolumn{5}{|c|}{ Herein } \\\hline
Baryon & quarks &  $(0,\tfrac{1}{2}^+)_3$   & $(1,\tfrac{1}{2}^+)_4$  &$(0,\tfrac{1}{2}^-)_5$   & $(0,\tfrac{1}{2}^+)_6$ & $(1,\tfrac{1}{2}^+)_7$ & $(0,\tfrac{1}{2}^-)_8$      & $(1,\tfrac{1}{2}^+)^\ast_9$       & $(0,\tfrac{1}{2}^-)^\ast_{10}$ \\\hline
$N        $       &   $uud$     &   0.938   &     1.440     &     1.535  &  0.948 &  1.279 &  1.144 &  1.440 &  1.542  \\
$\Lambda  $       &   $uds$     &   1.116   &     1.600     &     1.670  &  1.114 &  1.474 &  1.316 &  1.582 &  1.581  \\
$\Sigma   $       &   $uus$     &   1.189   &     1.660     &     1.620  &  1.114 &  1.474 &  1.316 &  1.582 &  1.581  \\
$\Xi      $       &   $uss$     &   1.315   &               &            &  1.279 &  1.670 &  1.487 &  1.723 &  1.620  \\\hline
$\Lambda_c$       &   $udc$     &   2.286   &               &     2.595  &  2.184 &  2.543 &  2.401 &  2.650 &  2.666  \\
$\Sigma_c $       &   $uuc$     &   2.455   &               &            &  2.184 &  2.543 &  2.401 &  2.650 &  2.666  \\
$\Lambda_b$       &   $udb$     &   5.619   &               &     5.912  &  5.394 &  5.809 &  5.650 &  5.916 &  5.916  \\
$\Sigma_b $       &   $uub$     &   5.811   &               &            &  5.394 &  5.809 &  5.650 &  5.916 &  5.916  \\\hline
$\Xi_c    $       &   $usc$     &   2.468   &               &     2.790  &  2.350 &  2.738 &  2.572 &  2.792 &  2.705  \\
$\Xi'_c   $       &   $usc$     &   2.577   &               &            &  2.350 &  2.738 &  2.572 &  2.792 &  2.705  \\
$\Xi_{cc} $       &   $ucc$     &   3.621   &               &            &  3.421 &  3.807 &  3.657 &  3.861 &  3.790  \\
$\Xi_b    $       &   $usb$     &   5.792   &               &            &  5.560 &  6.004 &  5.822 &  6.058 &  5.955  \\
$\Xi^\prime_b $   &   $usb$     &   5.945   &               &            &  5.560 &  6.004 &  5.822 &  6.058 &  5.955  \\
$\Xi_{cb}$        &   $ucb$     &           &               &            &  6.631 &  7.073 &  6.907 &  7.127 &  7.040  \\
$\Xi^\prime_{cb}$ &   $ucb$     &           &               &            &  6.631 &  7.073 &  6.907 &  7.127 &  7.040  \\
$\Xi_{bb} $       &   $ubb$     &           &               &            &  9.841 & 10.339 & 10.157 & 10.393 & 10.289  \\\hline
$\Omega_c $       &   $ssc$     &   2.695   &               &            &  2.516 &  2.934 &  2.744 &  2.934 &  2.744  \\
$\Omega_{cc} $    &   $scc$     &           &               &            &  3.586 &  4.002 &  3.829 &  4.002 &  3.829  \\
$\Omega_b $       &   $ssb$     &   6.046   &               &            &  5.726 &  6.200 &  5.994 &  6.200 &  5.994  \\
$\Omega_{cb} $    &   $scb$     &           &               &            &  6.796 &  7.268 &  7.079 &  7.268 &  7.079  \\
$\Omega_{ccb}$    &   $ccb$     &           &               &            &  7.867 &  8.337 &  8.164 &  8.337 &  8.164  \\
$\Omega_{bb} $    &   $sbb$     &           &               &            & 10.006 & 10.534 & 10.328 & 10.534 & 10.328  \\
$\Omega_{cbb}$    &   $cbb$     &           &               &            & 11.077 & 11.603 & 11.413 & 11.603 & 11.413  \\\hline
\end{tabular}
\end{center}
\end{table}

Given these observations, we turn now to computing the spectrum of $J=1/2^+\!$, $3/2^+$ baryon ground-states that contain one or more heavy quarks.  Following the approach used above for $u,d,s$ states, we first consider two theoretical constructs, \emph{viz}.\ $J^P=1/2^+$ nucleon-like states constituted from degenerate valence-quarks with $c$- or $b$-quark masses.  The nucleon Faddeev equation codes can be used to compute the masses of these states, with the results (in GeV):
\begin{subequations}
\begin{align}
m_{N_c} & = \phantom{1}4.66 \,,\; M_c^{(1/2) 0} = \tfrac{1}{3} m_{N_c} = 1.55\,,\\
%
m_{N_b} & =14.29 \,,\; M_b^{(1/2) 0} = \tfrac{1}{3} m_{N_b}  = 4.76\,.
\end{align}
\end{subequations}
Similarly, using the $J^P=3/2^+$ Faddeev equation, we obtain
\begin{subequations}
\label{MassesWQQQ}
\begin{align}
m_{\Omega_{ccc}} & = \phantom{1}4.76 \,,\;
M_c^{(3/2) 0} = \tfrac{1}{3} m_{\Omega_{ccc}} = 1.59\,,\\
m_{\Omega_{bbb}} & =14.37 \,,\;
M_b^{(3/2) 0} =\tfrac{1}{3} m_{\Omega_{bbb}} = 4.79\,.
\end{align}
\end{subequations}
These four values are reproduced in Table~\ref{SpectrumMq}.

\begin{figure}
\centering
\includegraphics[width=0.66\textwidth]{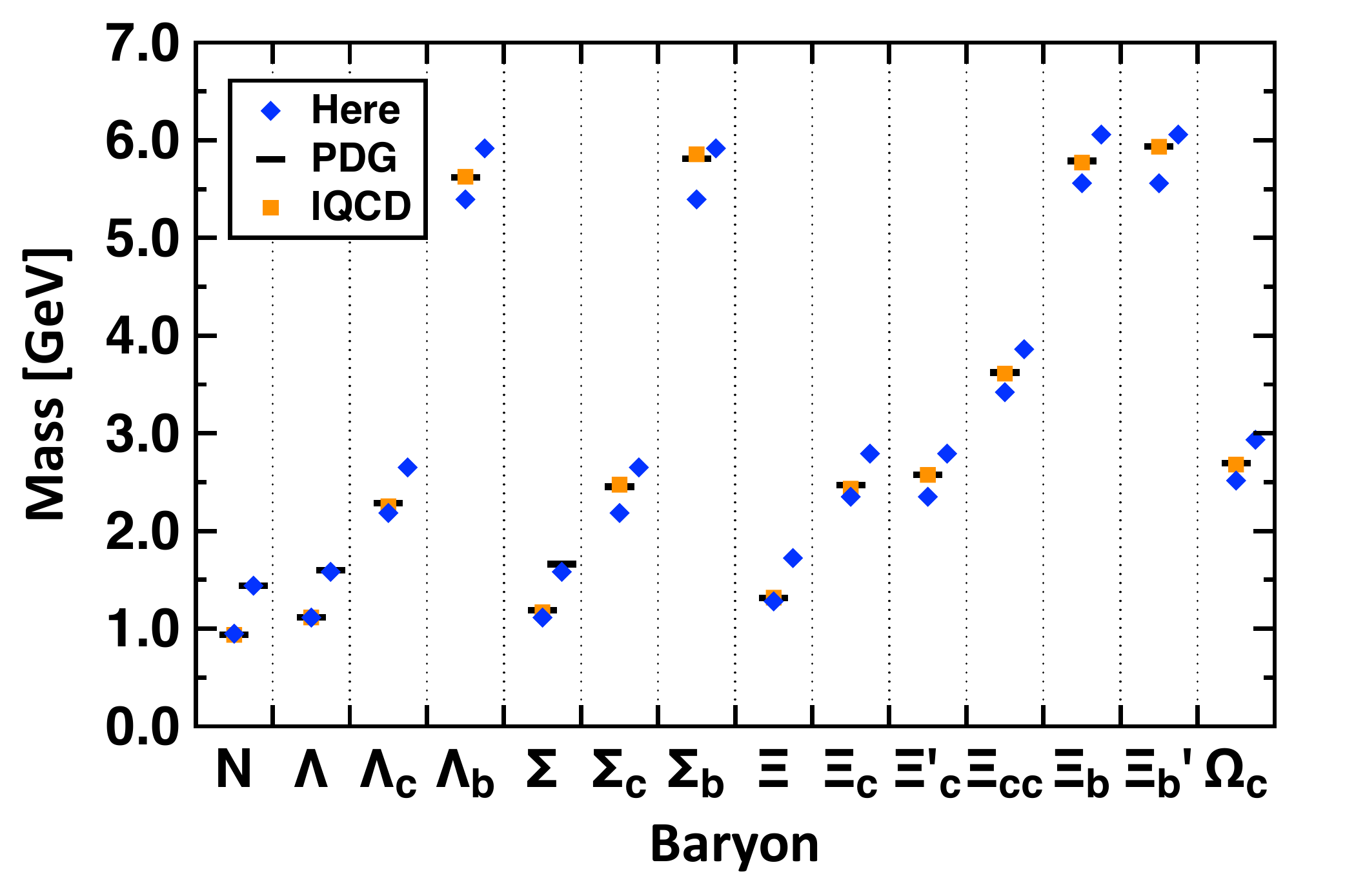} 
\includegraphics[width=0.66\textwidth]{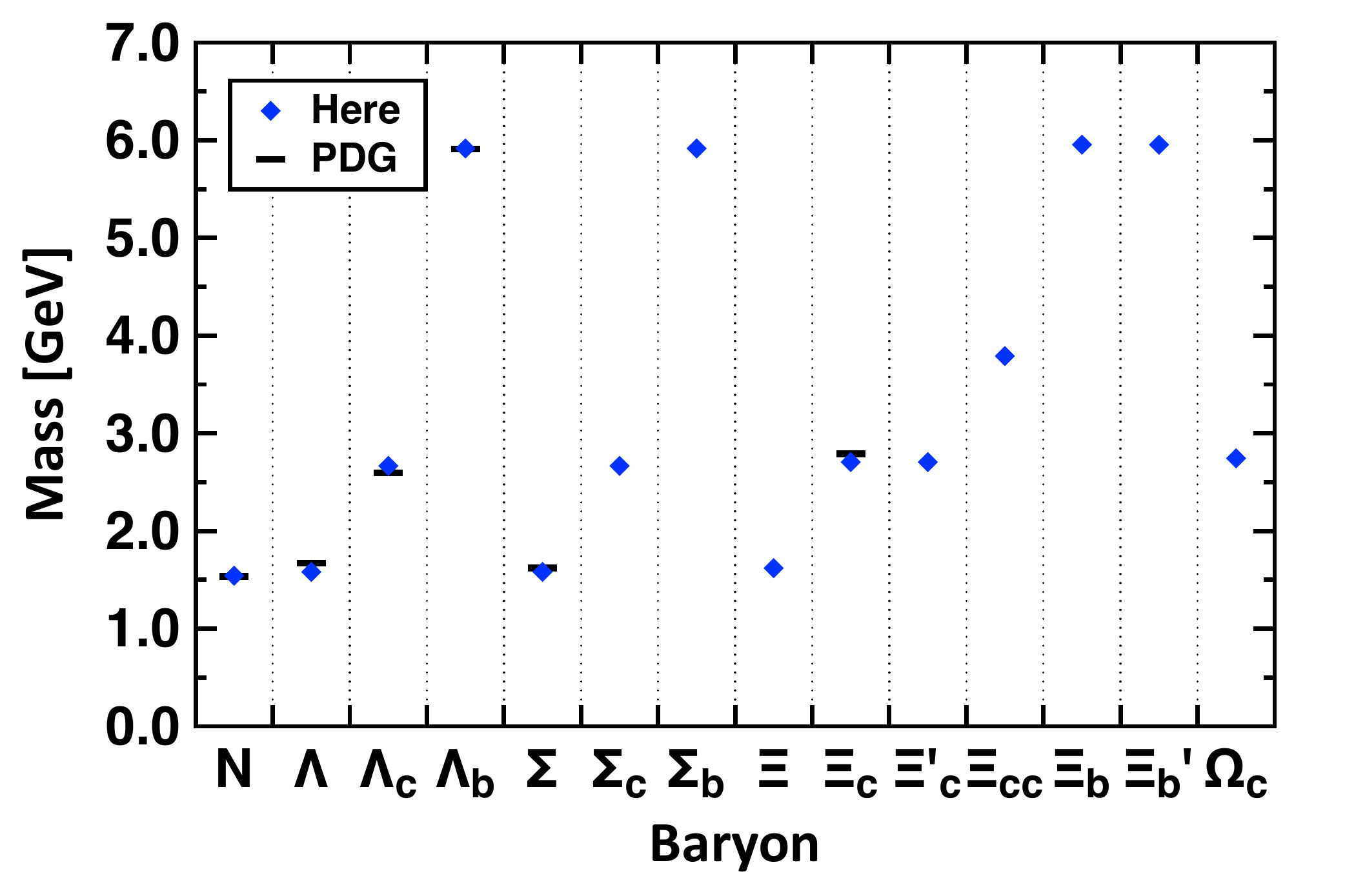} 
\caption{\label{JOneHalf}
Computed masses of selected $J=1/2$ baryons compared with experiment (PDG) \cite{Tanabashi:2018oca} and lQCD results \cite{Durr:2008zz, Brown:2014ena}.
\emph{Upper panel} -- positive parity ground states and their first positive parity excitations (displaced right); and \emph{lower panel} -- negative-parity ground states.
}
\end{figure}

\begin{figure}
\centering
\includegraphics[width=0.66\textwidth]{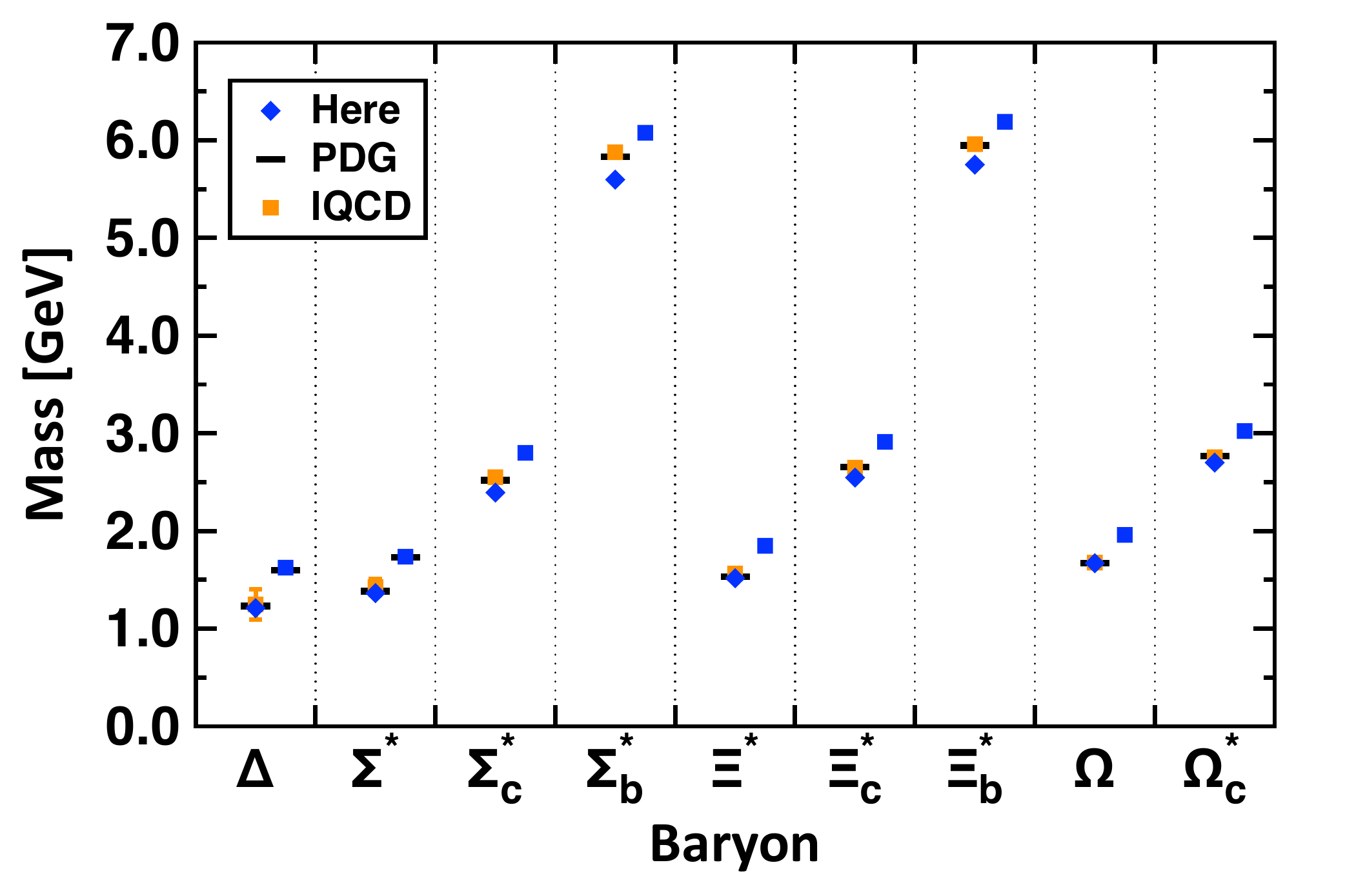} 
\includegraphics[width=0.66\textwidth]{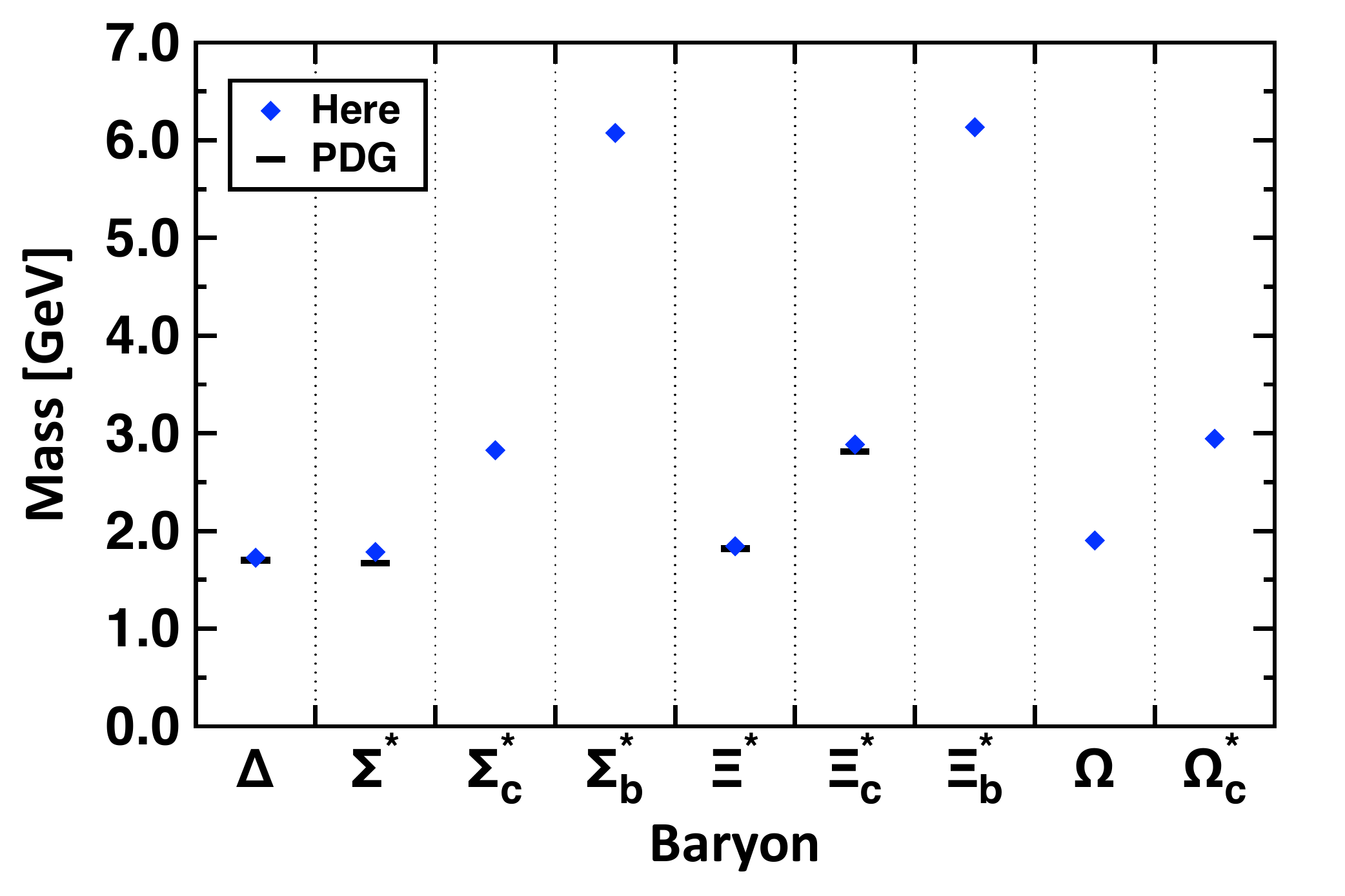} 
\caption{\label{JThreeHalf}
Computed masses of selected $J=3/2$ baryons compared with experiment (PDG) \cite{Tanabashi:2018oca} and lQCD results \cite{Durr:2008zz, Brown:2014ena}.
\emph{Upper panel} -- positive parity ground states and their first positive parity excitations (displaced right); and \emph{lower panel} -- negative-parity ground states.}
\end{figure}

Combining the results in Eqs.\,\eqref{OctetMq}\,--\,\eqref{MassesWQQQ} and using the ESRs described above, we obtain the masses of ground-state $J=1/2^+, 3/2^+$ $(q q^\prime q^{\prime\prime})$-baryons, where $q, q^\prime, q^{\prime\prime} \in\{ u,d,s,c,b\}$, listed in Tables~\ref{MassesTableOneHalf}, \ref{MassesTableThreeHalf}.
In addition, the upper panels of Figs.\,\ref{JOneHalf}, \ref{JThreeHalf} compare our predictions with empirical mass values in those cases for which that is possible.
The mean-absolute-relative-difference between the calculated values for the $(0,1/2^+)$ ground-states and the known empirical masses is $5.2(2.8) $\%; for the $(0,3/2^+)$ states, this difference is $2.6(1.6)$\%; and the combined difference is $4.2(2.7)$\%.
To provide context, considering the meson observables in Table~\ref{obsuds}, the agreement between RL truncation and experiment is $7.0(4.7)$\%.  Evidently, our level-based ESR procedure for computing the baryon \emph{spectrum} is no less reliable than its direct evaluation using RL truncation.

\begin{table}[!t]
\caption{\label{MassesTableThreeHalf}
Computed masses of $J=3/2$ baryons compared with experimental values \cite{Tanabashi:2018oca}, where known.  (Calculations assume isospin symmetry.)
States are labelled with the quark model name, drawn from Ref.\,\cite{Tanabashi:2018oca}, valence-quark content, and Faddeev equation identifications: $(n,1/2^P)$, where $n=0$ indicates channel ground-state and $n=1$ is the first like-parity excited state.  The numerical subscript indicates the table column number.
Columns labelled with an asterisk were computed as described in connection with Eqs.\,\eqref{sigmachangePP}, \eqref{sigmachangeRE}.
The mean-absolute-relative-difference between our best predictions (columns 6, 9, 10) and known experimental values is $3.6(2.7)$\%.
%
}
\begin{center}
\begin{tabular}{|c|c|l|l|l|r|r|r|r|r|}\hline
 \multicolumn{2}{|l}{\,} & \multicolumn{3}{|c}{ Empirical \cite{Tanabashi:2018oca} } & \multicolumn{5}{|c|}{ Herein } \\\hline
Baryon       & quarks &  $(0,\tfrac{3}{2}^+)_3$   & $(1,\tfrac{3}{2}^+)_4$  &$(0,\tfrac{3}{2}^-)_5$   & $(0,\tfrac{3}{2}^+)_6$ & $(1,\tfrac{3}{2}^+)_7$ & $(0,\tfrac{3}{2}^-)_8$      & $(1,\tfrac{3}{2}^+)^\ast_9$       & $(0,\tfrac{3}{2}^-)^\ast_{10}$ \\\hline
$\Delta     $     &   $uuu$     &   1.232   &  1.600   &   1.700  &  1.210 &  1.460 &  1.397 &  1.625 &  1.726  \\
$\Sigma^*   $     &   $uus$     &   1.383   &  1.730   &   1.670  &  1.363 &  1.627 &  1.565 &  1.737 &  1.785  \\
$\Xi^\ast      $     &   $uss$     &   1.532   &      	   &   1.820  &  1.517 &  1.793 &  1.734 &  1.848 &  1.843  \\
$\Omega     $     &   $sss$     &   1.672   &      	   &          &  1.670 &  1.960 &  1.902 &  1.960 &  1.902  \\\hline
$\Sigma^*_c $     &   $uuc$     &   2.518   &      	   &          &  2.393 &  2.690 &  2.607 &  2.800 &  2.826  \\
$\Xi^*_c    $     &   $usc$     &   2.656   &      	   &   2.815  &  2.547 &  2.857 &  2.775 &  2.912 &  2.885  \\
$\Xi^*_{cc} $     &   $ucc$     &           &      	   &          &  3.577 &  3.920 &  3.817 &  3.975 &  3.927  \\
$\Sigma^*_b $     &   $uub$     &   5.832   &      	   &          &  5.597 &  5.967 &  5.855 &  6.077 &  6.074  \\
$\Xi^*_b    $     &   $usb$     &   5.946   &      	   &          &  5.750 &  6.133 &  6.023 &  6.188 &  6.133  \\
$\Xi^*_{cb} $     &   $ucb$     &           &      	   &          &  6.780 &  7.197 &  7.065 &  7.252 &  7.175  \\
$\Xi^*_{bb} $     &   $ubb$     &           &      	   &          &  9.983 & 10.473 & 10.313 & 10.528 & 10.423  \\
\hline
$\Omega^*_c $     &   $ssc$     &   2.766   &      	   &          &  2.700 &  3.023 &  2.944 &  3.023 &  2.944  \\
$\Omega^*_{cc} $  &   $scc$     &           &      	   &          &  3.730 &  4.087 &  3.985 &  4.087 &  3.985  \\
$\Omega^*_{ccc} $ &   $ccc$     &           &      	   &          &  4.760 &  5.150 &  5.027 &  5.150 &  5.027  \\
$\Omega^*_b $     &   $ssb$     &           &      	   &          &  5.903 &  6.300 &  6.192 &  6.300 &  6.192  \\
$\Omega^*_{cb} $  &   $scb$     &           &      	   &          &  6.933 &  7.363 &  7.233 &  7.363 &  7.233  \\
$\Omega^*_{ccb} $ &   $ccb$     &           &      	   &          &  7.963 &  8.427 &  8.275 &  8.427 &  8.275  \\
$\Omega^*_{bb} $  &   $sbb$     &           &      	   &          & 10.137 & 10.640 & 10.481 & 10.640 & 10.481  \\
$\Omega^*_{cbb} $ &   $cbb$     &           &      	   &          & 11.167 & 11.703 & 11.523 & 11.703 & 11.523  \\
$\Omega^*_{bbb} $ &   $bbb$     &           &      	   &          & 14.370 & 14.980 & 14.771 & 14.980 & 14.771  \\\hline
\end{tabular}
\end{center}
\end{table}

It is worth remarking that within each of the pairs
$\Xi_c$-$\Xi^\prime_c$,
$\Xi_b$-$\Xi^\prime_b$,
$\Xi_{cb}$-$\Xi^\prime_{cb}$
in Table~\ref{MassesTableOneHalf} the members contain the same valence-quarks, respectively: $usc$, $usb$, $ucb$; and it is believed that in Nature each partner in the pair has these quarks arranged with different flavour symmetry, \emph{e.g}.\ $\Xi_c$ has $u\leftrightarrow s$ antisymmetry and $\Xi_c^\prime$ is symmetric under $u\leftrightarrow s$.
Herein, however, since the RL interaction kernel is flavour-blind, our analysis yields the same mass for each member of a given such pair.
On the other hand, if one employs a Faddeev kernel that expresses the appearance of diquark correlations within baryons, then flavour-symmetry is broken within the baryon wave functions and the flavour-antisymmetric state is always lighter than its flavour-symmetric partner because scalar diquarks are lighter than pseudovector diquarks.  As remarked following Eq.\,\eqref{MassN}, analogous effects contribute to the $\Lambda_{[ud]s}^0$-$\Sigma_{\{ud\}s}^0$ mass splitting.

\subsection{Parity Partners in the Baryon Spectrum}
\label{ParityPartners}
All Poincar\'e-covariant studies of the hadron spectrum predict opposite-parity partners of a given ground-state; and in relativistic quantum field theory, one may generate the interpolating field for the parity partner via a chiral rotation of that associated with the original state.  It follows that parity partners will be degenerate in mass and alike in structure in all theories that possess a chiral symmetry realised in the Wigner-Weyl mode.  (There is evidence of this, \emph{e.g}.\ in both continuum \cite{Maris:2000ig, Wang:2013wk} and lattice \cite{Cheng:2010fe, Aarts:2017rrl} analyses that explore the evolution of hadron properties with temperature.)
Such knowledge has long made the mass-splittings between parity partners in the strong-interaction spectrum a subject of interest.

A well-known example is that provided by the $\rho(770)$- and $a_1(1260)$-mesons: viewed as chiral and hence parity partners, it has been argued \cite{Weinberg:1967kj} that their mass and structural differences can be attributed entirely to DCSB, \emph{viz}.\ realisation of chiral symmetry in the Nambu-Goldstone mode.  It is plausible that this profound emergent feature of the Standard Model is tightly linked with confinement \cite{Roberts:2016vyn}; and regarding DCSB's role in explaining the splitting between parity partners, additional insights have been developed by studying the bound-state equations appropriate to the $\rho$- and $a_1$-mesons.  In their rest frames, one finds that their Poincar\'e-covariant wave functions are chiefly $S$-wave in nature \cite{Maris:1999nt, Chang:2008sp, Chang:2011ei, Roberts:2011cf, Chen:2012qr, Eichmann:2016yit}, even though both possess nonzero angular momentum \cite{Bloch:1999vka, Gao:2014bca}, whose magnitude influences the size of the splitting \cite{Chang:2011ei}.

It only became possible to elucidate the impact of orbital angular momentum on hadron masses and, hence, reliably treat parity partners in the meson spectrum after techniques were developed that enable DCSB to be expressed in the Bethe-Salpeter kernel \cite{Chang:2009zb, Chang:2011ei, Binosi:2014aea, Qin:2016fwx, Binosi:2016rxz, Williams:2015cvx}.  Likewise, as evident in the comparison between columns 5 and 8 in Tables~\ref{MassesTableOneHalf}, \ref{MassesTableThreeHalf}, the RL truncation is unable to explain the splittings between parity partners in the baryon spectrum.  As with kindred mesons, what lacks is DCSB-enhanced repulsion involving $P$-wave components of the hadron wave-functions.  An efficacious phenomenological remedy was proposed in Ref.\,\cite{Roberts:2011cf} and has since been used elsewhere \cite{Chen:2012qr, Eichmann:2016yit}.  Namely, the effects of DCSB-induced repulsion in the kernels of the Bethe-Salpeter and Faddeev equations for negative-parity channels should be mimicked by suppressing the strength of the exchange-interaction between light quarks.  Hence, the entries in columns 10 of Tables~\ref{MassesTableOneHalf}, \ref{MassesTableThreeHalf} were obtained by modifying the interaction as follows: if, and only if, the interaction takes place between two light quarks, then
\begin{equation}
\label{sigmachangePP}
\varsigma \to \varsigma_l^- = 0.85 \, \varsigma
\end{equation}
in Eq.\,\eqref{varsigmalight}.  The interaction strength is unchanged if one or both the quarks involved is $s$, $c$, $b$.
The computed spectrum-constituent masses appropriate for these states are listed in the $(0,1/2^-)^\ast$ and $(0,3/2^-)^\ast$  columns of Table~\ref{SpectrumMq}.  This procedure improves the mean-absolute-relative-difference between our computed results for the negative-parity ground-states and the associated experimental masses by a factor of five, \emph{viz}.\ 14(9)\% $\rightarrow$ 3(2)\%.

In constituent-quark potential models it is usual to describe the lightest negative-parity partners of ground-state baryons as $P$-wave states \cite{Isgur:1978xj}, \emph{viz}.\ quantum mechanical systems with one unit of constituent-quark orbital angular momentum, $L$, coupled with the constituent-quark spin, $S$, to form the total angular momentum of the bound-state: $J=L+S$.  In relativistic quantum field theory, however, $L$ and $S$ are not good quantum numbers.  Moreover, even if they were, owing to the loss of particle number conservation, it is not clear \emph{a priori} just with which degrees-of-freedom $L$, $S$ should be connected.  This question is related to the fact that the constituent-quarks used in building quantum mechanical models have no known mathematical connection with the degrees-of-freedom featuring in QCD.  Notwithstanding these issues, one typically finds \cite{Eichmann:2016nsu, Chen:2017pse, Qin:2018dqp, Chen:2019fznFBS}, at least for the lower-lying states, some support in quantum field theory for the constituent-quark model classifications of such systems when using Faddeev equations of the type depicted in Fig.\,\ref{FEimage}, which describe baryon structure and dynamics at a typical hadronic scale in terms of dressed-quark degrees-of-freedom.
Hence, there is a sense in which dressed-quarks, whose properties can be and are calculated in QCD, serve as Nature's embodiment of the constituent-quarks used so effectively in beginning to bring order to hadron physics \cite{GellMann:1964nj, Zweig:1981pdFBS}.

\subsection{Positive-Parity Excitations of the Ground-State Baryons}
Ever since discovery of the proton's first positive-parity excitation, the Roper resonance \cite{Roper:1964zza, BAREYRE1964137, AUVIL196476, PhysRevLett.13.555, PhysRev.138.B190}, there have been questions concerning the character of like-parity excitations of ground-state positive-parity baryons.  Now, a coherent picture is emerging following \cite{Burkert:2017djoFBS}:
(\emph{i}) the acquisition and analysis of a vast amount of high-precision nucleon-resonance electroproduction data with single- and double-pion final states on a large kinematic domain of energy and momentum-transfer;
(\emph{ii}) development of a sophisticated dynamical reaction theory capable of simultaneously describing all partial waves extracted from available, reliable data;
(\emph{iii}) formulation and wide-ranging application of a Poincar\'e covariant approach to the continuum bound state problem in relativistic quantum field theory that expresses diverse local and global impacts of DCSB in QCD;
and \emph{(iv}) the refinement of constituent quark models so that they, too, qualitatively incorporate these aspects of strong QCD.
In this picture such states are, at heart, radial excitations of the associated ground-state baryon, consisting of a well-defined dressed-quark core, augmented by a meson cloud.

As remarked above, in choosing the scale in Eq.\,\eqref{varsigmalight} so as to describe a given set of light-hadron observables using RL truncation, some influences of the meson cloud are implicitly incorporated.  Important features are still omitted, however; \emph{e.g}.\ baryon resonances studied in RL truncation do not have widths, which are an essential physical consequence of meson-baryon final-state interactions (MB\,FSIs).  The operating conjecture for RL truncation is that the impact of MB\,FSIs on a resonance's Breit-Wigner mass is captured by the choice of interaction scale, even though a width is not generated.  This should be reasonable for states whose width is a small fraction of their mass; and in practice, as already illustrated herein and in many other studies, the conjecture appears to be correct, at least for the ground-state $J=1/2^+$, $3/2^+$ systems.

Turning to the first positive-parity (radial) excitations of hadrons, RL truncation is known to be deficient in some other ways, \emph{e.g}.\ in the meson sector it typically produces excited states that are too light \cite{Holl:2004fr} and potentially ordered incorrectly \cite{Qin:2011xq}.
Regarding Tables~\ref{MassesTableOneHalf}, \ref{MassesTableThreeHalf}, it is evident that the masses of positive-parity excitations of ground-state light-quark baryons are also underestimated by RL truncation.

We highlighted in Sec.\,\ref{ParityPartners} that when a bound-state calculation underestimates the mass of a given state, the most obvious culprit is an interaction kernel providing too much attraction or, equivalently, too little repulsion.   Therefore, following the success of the rescaling in Eq.\,\eqref{sigmachangePP} for negative-parity baryons, we checked whether a similar expedient can also be effective for the first positive-parity excitations.  The results in column~9 of Tables~\ref{MassesTableOneHalf}, \ref{MassesTableThreeHalf} were obtained with
\begin{equation}
\label{sigmachangeRE}
\varsigma \to \varsigma_l^R = 0.93 \, \varsigma
\end{equation}
in Eq.\,\eqref{varsigmalight}.  Again, the interaction strength is unchanged if one or both of the quarks involved is $s$, $c$, $b$.\footnote{Given current experimental data on the splittings between parity partners and radial excitations in systems with heavier quarks, one cannot be certain whether the interaction strength should be changed in $s$, $c$, $b$ channels.  Theoretically, on the other hand, if these observed splittings are driven by DCSB, as we believe, then the effects should diminish with increasing current-quark mass.  In that case, within the accuracy of our approach, it is sensible to modify only the light-quark interaction strength.}
This procedure improves the mean-absolute-relative-difference between our computed results for the radial excitations of the positive-parity ground-states and the associated experimental masses by a factor of six, \emph{viz}.\ 9.0(2.2)\% $\rightarrow$ 1.6(1.9)\%.
Moreover, the correction brings the masses predicted for the positive-parity excitations of the $\Xi$, $\Xi^\ast$, $\Omega$ baryons, empirically ``missing'' from the octet and decuplet, into line with those inferred from the Poincar\'e-covariant quark-diquark Faddeev equation analysis in Ref.\,\cite{Chen:2019fznFBS}, \emph{viz}.\ $m_\Xi=1.75(12)$, $m_{\Xi^\ast}=1.89(03)$, $m_{\Omega}=2.05(02)$.
%

Considering the flavour-diagonal systems computed directly herein, we have checked all components of the associated rest-frame-projected Poincar\'e-covariant wave functions and found that for any given component there is always at least one kinematic configuration for which it exhibits a single zero.  There are no configurations for which any amplitude possesses more than one zero.  (Ref.\,\cite{Qin:2018dqp}, Sec.\,IV.C, provides further details.)

Drawing upon experience with quantum mechanics and studies of excited-state mesons using the Bethe-Salpeter equation \cite{Holl:2004fr, Qin:2011xq, Li:2016dzv, Li:2016mah},  such features are indicative of a first radial excitation.
Notwithstanding that, given the complexity of Poincar\'e-covariant wave functions for baryons, shifts in the relative strengths of various angular-momentum components are usually also found within the wave function of a baryon's like-parity excitation \cite{Eichmann:2016nsu, Chen:2017pse, Qin:2018dqp, Chen:2019fznFBS}.

\section{Epilogue}
\label{epilogue}
Using a symmetry-preserving rainbow-ladder truncation of the appropriate bound-state equations in relativistic quantum field theory, with particular emphasis on the Poincar\'e-covariant Faddeev equation, we described a calculation of the spectrum of ground-state $J=1/2^+$, $3/2^+$ $(qq^\prime q^{\prime\prime})$-baryons, where $q, q^\prime, q^{\prime\prime} \in\{ u,d,s,c,b\}$, their first positive-parity excitations and parity partners.  Employing two parameters, one relating to the interaction strength in the parity-partner channels and the other to that in positive-parity excitations, our analysis reproduces the known spectrum of 39 such states with an accuracy of $3.6(2.7)$\%.  Where our predictions drift from the empirical values, they are systematically below the known mass owing to deficiencies in RL truncation whose origin is understood.  From this foundation, we proceeded to predict the
masses of 90 states not yet seen empirically.

Our approach also yields the Poincar\'e-covariant wave functions for many of these states; and whilst we did not scrutinise their properties herein, it will be worth doing so in future.  Existing analyses of this type have provided insights that, \emph{e.g}.\ reveal which of those structural perspectives provided by constituent-quark potential models are qualitatively robust, and also enrich the understanding of all these systems.
Furthermore, with wave functions in hand, one can also compute an array of dynamical observables, including, \emph{inter alia}: electroweak couplings and form factors; and strong transition form factors.  Such quantities provide connections with observables that are particularly sensitive to the internal structure of these basic yet complex strong-interaction bound-states.

Finally, so far as continuum bound-state studies are concerned, no material improvements over the analysis and results described herein can be envisaged before
the general spectral function methods introduced elsewhere \cite{Chang:2013pq} for meson bound-state problems have been extended to baryons
and/or the relevant interaction kernels are improved, to incorporate nonperturbative effects of dynamical chiral symmetry breaking and express measurable effects of resonant contributions.
Such efforts are likely to benefit from the use of high-performance computing.

\begin{acknowledgements}
We are grateful for constructive comments and encouragement from
L.~Chang,
C.~Chen,
Z.-F.~Cui,
R.~Gothe,
V.~Mokeev,
J.~Segovia,
S.-S.~Xu
and
P.-L.~Yin;
and for the hospitality of RWTH Aachen University, III.\,Physikalisches Institut B, Aachen, Germany.
Work supported by:
National Natural Science Foundation of China (NSFC) under Grant No.\,11805024;
%
%
Jiangsu Province \emph{Hundred Talents Plan for Professionals};
U.S.\ Department of Energy, Office of Science, Office of Nuclear Physics, under contract no.~DE-AC02-06CH11357;
and
Forschungszentrum J\"ulich GmbH.
\end{acknowledgements}



\end{document}